\documentstyle[12pt]{article}
\begin{document}
\pagestyle{empty}
\begin{center}
{\large {\bf A  renormalization group analysis of extended \\
electronic states in 1d quasiperiodic lattices}}  \\
\vskip 1cm
Arunava Chakrabarti\footnote{{\bf Permanent  Address}:  Scottish  Church
College,
1\&3 Urquhart Square, Calcutta 700~006, India}, S. N. Karmakar and R. K.
Moitra\\
{\em Saha Institute of Nuclear Physics, 1/AF, Bidhannagar, Calcutta
700 064,
India.}\\
\vskip 1 cm
{\bf ABSTRACT}
\end{center}
We  present   a   detailed   analysis   of the   nature  of electronic
eigenfunctions in one-dimensional quasi-periodic  chains based on a
clustering idea recently introduced by us [Sil et al., Phys. Rev. {\bf B
48}, 4192
(1993) ], within   the   framework   of   the   real-space   renormalization
group
approach. It is shown that even in the absence of translational  invariance,
extended states arise in a class of such lattices if they possess a  certain
local correlation among the constituent atoms. We have applied  these  ideas
to the quasi-periodic period-doubling chain,  whose  spectrum  is  found  to
exhibit a rich variety of behaviour, including a cross-over from  critical
to an extended  regime,  as  a  function  of  the  hamiltonian  parameters.
Contrary to prevailing  ideas,  the  period-doubling  lattice  is  shown  to
support  an  infinity  of  extended  states,  even  though  the   polynomial
invariant
associated with the trace map is non-vanishing. Results  are  presented  for
different  parameter  regimes,  yielding  both  periodic  as  well   as
non-periodic eigenfunctions. We have also extended the present theory  to  a
multi-band  model  arising  from  a  quasi-periodically  arranged  array  of
$\delta$-function potentials on the atomic  sites.  Finally,  we  present  a
multifractal  analysis  of  these  wavefunctions  following  the  method  of
Godreche and Luck [ C. Godreche and J. M. Luck, J. Phys. A :Math.  Gen. {\bf
23},
3769 (1990)] to confirm their extended character.
\newpage
\pagestyle{plain}
\setcounter{page}{1}
\noindent
{\bf I.  Introduction}

      In the last decade there has  been  intense  activity  in  theoretical
studies of aperiodic lattices in one dimension. Apart from  their  intrinsic
interest, these  studies  have  been  mainly  motivated  by  the  remarkable
discovery of the so-called quasicrystalline ordering  in  solids,  beginning
with the work of Shechtman et al.$^1$ in 1984. On the experimental side, one
dimensional aperiodic lattices  have  also  become  interesting  objects  of
study,  following  the  successful  fabrication  of  layered  semiconducting
superstructures grown epitaxially  in  accordance  with  the  rules  of  the
well-known  Thue-Morse  and  Fibonacci  sequences$^2$.  Parallel  with  this
development, theoretical  studies  have  been  made  on  a  number  of  such
one-dimensional systems for phonon, electron and magnon  spectra$^3$.  While
most of these studies have been numerical, there have also been quite a  few
analytical attempts,  starting  with  the  pioneering  work  of  Kohmoto  et
al.$^4$
on the unusual electronic properties of the Fibonacci lattice.

               The   behaviour   of   the    electronic    eigenstates    in
one-dimensional lattices are well-known in the two extreme limits of  random
and perfect periodic ordering; in the former case all  the  eigenstates  are
localised, while in the latter they are extended. It  is  precisely  because
aperiodicity is a level of ordering intermediate between these  two  extreme
limits that the electronic eigenstates in aperiodic linear systems often are
found  to  be  critical$^4$.  Interestingly,  there  is  also  evidence  for
extended states in some of these systems$^{5-9}$. While the reason  for  the
existence of critical states may be assigned to the lack of  periodicity  in
these systems, extended states arise due  to  other  reasons,  as  discussed
below.

           In a recent work$^5$ we analysed this  problem in one-dimensional
systems and found that the basic reason may be traced to the existence of  a
particular type of correlation among the atoms in the lattice. For the  sake
of clarity we summarise here the main ideas developed in that work.  In   an
elementary problem of Anderson localisation of the  random  distribution  of
$A$ atoms on a host lattice of $B$ atoms, it is well-known that there is  an
extended eigenstate at an energy $\epsilon_A$ in  this  system  if  the  $A$
atoms always occur in pairs, that is as dimers. This is  an  instance  of  a
definite kind of correlation leading to an extended state  in  a  disordered
system$^6$. In this case the correlation consists in the fact  that  an  $A$
atom always appears as a member of a dimer. We may generalise this  idea  as
follows. Consider the distribution of  a  certain  well-defined  cluster  of
atoms on a host lattice in one dimension. Suppose that this cluster consists
of a finite number  of identical building blocks , each of which  is  a
smaller unit consisting, in general, of several atoms. In the above example,
for instance, a dimer may be regarded as a cluster of  two  consecutive  $A$
atoms. If the building blocks are to be found only within the  clusters  and
nowhere else on the lattice, then it can be shown that  for  certain  energy
values the contribution to the total transfer  matrix  for  the  chain  from
these clusters is unity. Therefore, at these special energies only the  host
lattice is effective in determining  the electronic properties. If the  host
lattice is periodic, then there is a possibility of extended states at these
energies for the whole sytem.

          The special energies for which the   states  are  extended  are  a
property of the cluster only and do not depend on the lattice  as  a  whole.
Thus in our earlier example of random  $AA$  dimers   on  a  $B$  type  host
lattice , the  special  energy  value  $E  =  \epsilon_A$    is   determined
by
considering the $AA$ cluster alone.  Many  other  $1d$  lattices  containing
these $AA$ dimers may have  an  extended  state  at  this  same  energy.  To
illustrate this let us consider a periodic lattice of $A$  atoms.  Here  the
$AA$ correlation is trivially present, and  therefore  this  system  has  an
extended state at  the  same  energy.  However,  while  all  eigenstates  at
energies other than $\epsilon_A$ are localised in  the  random  dimer  case,
here  we have a whole band of extended states, the existence of which cannot
be inferred by considering the $AA$ dimers. Although Bloch's theorem enables
us to find this band directly, we may  alternatively  think  of  determining
this band by considering  higher  order  clusters  consisting  of  triplets,
quadruplets  \ldots  etc.  of  $A$  atoms,  as  has  been  discussed  by  us
earlier$^5$. There is no advantage in adopting this point of  view  in  this
elementary example. However, in aperiodic chains Bloch's  theorem  does  not
apply, and one has to use this idea in  order  to  determine  the  bands  or
mini-bands of extended states.

               The clusters  contain  only  partial  information  about  the
entire lattice, and consequently, we have the same set of energy levels  for
systems all of which contain a particular cluster but which  have  different
long-range compositions. The detailed differences in the electronic  spectra
in these systems are due to correlations at length scales beyond the cluster
size.  In  order  to  bring  out  these  differences  we  have  to   examine
correlations among larger and larger  blocks  of  atoms  in  the  chain,  by
looking at the systems at increasing  length scales.  In  this  process  the
long range features of the lattice get gradually included  and we end up  by
obtaining  the  entire  spectrum  of  extended  states.   The   real   space
renormalization group (RSRG) method is particularly suited for studying this
kind of problem.

                    In  this  paper  we  consider  a  period-doubling   (PD)
chain$^{10,12}$ as a prototype example to illustrate  the  above  clustering
ideas. These ideas are quite  general  and  are  also  applicable  to  other
aperiodic systems.   A  PD  chain,  to  our  mind,  is  a  very  interesting
candidate for investigation beacuse the trace map associated with this chain
leads to a polynomial invariant which  is  likely  to  ensure  the  critical
nature of all the eigenstates as well as a Cantor set energy spectrum. As we
will show in this paper, inspite of the existence of such an invariant, this
lattice {\em can} sustain an infinity of extended eigenstates which  coexist
with
the otherwise Cantor set nature of the energy spectrum.

           The few analytical studies that are available on the  PD  lattice
have all been made with a  simple   on-site  tight-binding  version  of  the
lattice hamiltonian.   Bellissard et al.$^{11}$ have shown that the spectrum
of a  tight  binding  Schrodinger  equation  in  which  the  potentials  are
distributed following a  PD  sequence  is  purely  singular  continuous  and
supported on a Cantor set of zero Lebesgue measure. In  an  earlier  article
Severin et al.$^{12}$  analytically studied the case of a special  class  of
quasiperiodic sequences  generated  by  the  substitution  rule  $S_{L+1}  =
S_{L-1}S_{L-1}S_{L}$ which yielded the  standard  copper  mean  lattice  and
variations of PD lattices on tuning the initial choice of the basic building
blocks. They reported the existence of periodic eigenstates on such lattices
at some special values of energy.  From our  point  of  view,  the  work  of
Severin et al. is particularly  interesting,  as  they  give  an  analytical
method  for  locating  the  eigenvalues  and  the   corresponding   extended
eigenstates for the PD sequence. The validity of their approach, however, is
strictly limited to the use of the on-site model.  Apart  from  this,  their
analysis  crucially depends  on  a  special  choice  of  the  value  of  the
pseudo-invariant$^{12}$ $\gamma$ for PD chain equal to 2, and it is not  clear
how
this analysis can be extended to other values of $\gamma$.

           Interestingly, in all the works on the PD chain reported so  far,
the clustering effect that we have discussed above,  although  present,  has
neither received adequate emphasis, nor has been made the  focal  point  for
analysing the nature of the eigenfunctions.  As will be seen, the clustering
idea enables us to reduce the problem of identification  of  the  nature  of
eigenfunctions on any quasi-periodic lattice to  its  essential  simplicity.
Thus, for instance, by merely looking at the Fibonacci chain it is  possible
to infer that there are no extended states in general,  without  going  into
any further analytical consideration, since there is no clustering  in  this
chain$^8$.

          One of the aims of this paper is to  investigate  the  nature  and
character of the electronic eigenfunctions  with more general hamiltonians.
Such an  extension  is necessary  because any experimental realisation of  a
quasi-periodic chain  will  not  be  restricted  to  on-site  variations  of
parameters only. As a first step we therefore look at a tight-binding  model
hamiltonian for the PD chain  which includes simultaneous variations in both
on-site and  hopping  matrix  elements.   This  increase  in  the  level  of
complexity is not readily amenable to analysis through the  transfer  matrix
approach alone. For the PD case  we find interesting cross-over behaviour in
the nature of the eigenstates from an {\em all states critical} picture to a
situation where we have {\em an infinity of  extended   states},   depending
on
the
region of parameter space in which we work.  We next  consider the   problem
of the motion of an electron in  an array of  $\delta$- function  potentials
whose  strengths  are  distributed  in  accordance   with   the   underlying
quasi-periodicity of the lattice. This is a multi-band problem, and we  find
that the clustering idea  still  applies.  We  may  regard  this  case  as
approximating the experimental situation more closely than the  single  band
model based on the tight binding approach.

          We organise this paper as follows. In the next section we  give  a
somewhat detailed exposition of  the  clustering  idea  in  the  context  of
general single band tight binding models of a  PD  chain.  We  also  present
numerical results for the extended eigenfunctions obtained on the  basis  of
the present theory. The extension to the multiband case is given in  section
III.  Section  IV  deals  with  the
multifractal analysis of the wave functions.

\vskip .5cm
\noindent
{\bf II. The Tight Binding Model}

          A portion of the period doubling chain  is  shown   in   Fig.1(a).
The
sequence in which the long ($L$) and short ($S$) bonds are arranged in  this
chain is  obtained  by  successively  using  the  substitution   rule$^{10}$
$L\rightarrow
LS$ and $S\rightarrow LL$.  For  describing  the  electron  states  in  this
lattice we use the single band tight binding hamiltonian
\begin{equation}
          H= \sum_i \epsilon_i |i><i| + \sum_{<ij>} t_{ij} |i><j| .
\end{equation}

          Since  we  are  interested  in  locating  clustering  effects   at
different length scales, it is essential to label the sites and the  hopping
matrix elements appropriately. This  is  necessary,  because  in  order  to
implement the RSRG decimation scheme, the self-similarity inherent  in  the
lattice has  to  be  preserved  at  every  stage  of  renormalization.  Such
labelling was discussed by us earlier for  the  Fibonacci  lattice  and  the
copper  mean  lattice$^{5,8}$.  There  is  a   basic   difference   in   the
renormalization procedure followed in the transfer matrix method and that in
the present approach.  Our  approach  is  based  on  the  global  decimation
procedure discussed by Southern et  al.$^{14}$,  in  which  we  consider  an
infinitely long chain, and decimate a subset of sites so  as  to  produce  a
self-similar chain on an inflated length scale. We  find  that  we  have  to
assign at least three different labels to the site energies and two  to  the
hopping matrix elements to make this decimation possible in  the  PD  chain.
Consequently we identify three types of sites, which we label  by  $\alpha$,
$\beta$  and  $\gamma$  corresponding  respectively  to  sites  between  two
consecutive long ($L$) bonds, a long ($L$) bond and a short ($S$) bond,  and
a short ($S$) bond and a long  ($L$)  bond.  The  site  energies  in  Eq.(1)
therefore  assume  the  values   $\epsilon_\alpha$,   $\epsilon_\beta$   and
$\epsilon_\gamma$ respectively. The hopping matrix elements across the  long
bond ($L$) is taken as $t_L$ and that across the short bond ($S$) as  $t_S$.
As we shall see, the site energies and the hopping matrix  elements  of  the
renormalized chain at any stage carry these very five labels and no more. Of
course later we shall discuss the very interesting  case  of  a  still  more
general model which has four types of hopping  matrix  elements  across  the
long ($L$) bond depending on the vertices connected, and whose behaviour  is
quite different requiring  a separate treatment.

          Let us begin with the case  where  we  have  three  site  energies
$\epsilon_\alpha$, $\epsilon_\beta$ and $\epsilon_\gamma$  and  two  hopping
matrix elements $t_L$ and $t_S$. The eigenfunctions for such a  lattice  may
be calculated  by  the  standard  transfer-matrix  method.  The  Schrodinger
equation for the hamiltonian (1) can be cast in the  form  $\phi_{n+1}=  M_n
\phi_n$, where
\[
     \phi_n = \left (\begin{array}{c}
               \psi_n \\ \psi_{n-1}
\end{array}
\right )\ \mbox{\rm and } M_n = \left( \begin{array}{cc}
\frac{E-\epsilon_n}{t_{n,n+1}} & -\frac{t_{n,n-1}}{t_{n,n+1}} \\
1 & 0\end{array}\right).
\]
Here $\psi_n$ denotes the amplitude of the wave  function  at  the
$n$-th site and $M_n$ is a $2\times 2$ transfer matrix. In the PD  chain  we
have three different kinds of transfer matrices  $M_\alpha$,  $M_\beta$  and
$M_\gamma$, and $\phi_n$ is related to $\phi_0$ by a product of these  three
matrices following the PD sequence.

           By inspecting  the  PD  chain  (Fig.1(a))   we   see   that   the
$\alpha
$-sites {\em always} occur in pairs whereas, the $\beta$ and $\gamma$  sites
always occur as a doublet $\beta \gamma$. Thus for this lattice the pair  of
sites $\alpha -\alpha$ constitutes a cluster in the sense discussed  in  the
Introduction. For this chain the string of transfer matrices typically looks
like$  \ldots   M_{\beta   \gamma}.M_\alpha^2.   M_{\beta   \gamma}.M_{\beta
\gamma}.M_{\beta  \gamma}.M_\alpha^2 \ldots$,   where   $M_{\beta    \gamma}
=M_\gamma
M_\beta$. We notice that the matrix $M_\alpha  $  is  unimodular.  The  pure
on-site model results as  a  special  case  by  taking  $\epsilon_\alpha   =
\epsilon_\gamma  = \epsilon_A$, $\epsilon_\beta  =  \epsilon_B$   and   $t_L
=
t_S$ while the transfer model is  obtained  by  taking  $\epsilon_\alpha   =
\epsilon_\beta =\epsilon_\gamma =\epsilon$ and $t_L\neq t_S$.

                Let us first discuss the general case of a  one  dimensional
chain  which  contains  $m$-component  clusters,  that  is   clusters   each
containing $m$ identical building blocks. Let the transfer matrix for  every
building block be a $2 \times 2$  unimodular  matrix  denoted  by  $M$.  The
composite  transfer  matrix  for  the  cluster  is   then   $M^m$.    By   a
straightforward application of the Cayley-Hamilton theorem  for  the  $m$-th
power of a $2 \times 2$ unimodular matrix $M$ it can be shown that
\begin{equation}
M^m = U_{m-1}(x)M - U_{m-2}(x)I
\end{equation}
where, $x= (1/2) {\rm Tr}M$. $U_m(x) = \sin (m+1)\theta /\sin\theta $,  with
$\theta = \cos^{-1}x$  is the $m$-th  Chebyshev  polynomial  of  the  second
kind$^5$.   For a value of the energy $E$  for  which  $U_{m-1}(x)$  becomes
zero,  $M^m$ reduces to $-U_{m-2}  (x)I$,  that  is  to  say,  the  transfer
matrix
for this cluster behaves essentially as the identity matrix at this  energy.
Thus at this energy the entire lattice does not feel  the  presence  of  the
cluster defined by the transfer  matrix  $M^m$.  If  the  remainder  of  the
lattice forms a periodic chain, there will be extended states at this energy
provided, this energy is an allowed one. For allowed states  wave  functions
do not diverge at infinity.

          For the PD chain  $M = M_\alpha $ and $m=2$ and therefore  setting
$U_1(x)=0$ we obtain $E=\epsilon_\alpha $. For this energy the whole lattice
effectively behaves as a periodic diatomic linear chain with each unit  cell
containing  a  $\beta  $  and  a  $\gamma$  atom.  If  the   energy   $E   =
\epsilon_\alpha$ happens to be within the  allowed  band  of  this  diatomic
lattice, then this energy is an allowed  one.  The  condition  for  this  is
$(1/2) |\mbox{\rm Tr}M_{\beta \gamma}|\leq 1$.

          We now proceed to determine other  energy  eigenvalues  which  may
give rise  to  extended  eigenfunctions  for  the  entire  chain.  From  the
Schrodinger equation for the PD chain (see Fig.1(a)) we obtain the following
hierarchy of equations for  the
amplitudes of the wavefunction
\begin{eqnarray}
\vdots \nonumber \\
     (E-\epsilon_\gamma)\psi_{-1}& = & t_L \psi_0 +t_S  \psi_{-2}  \nonumber
\\
     (E-\epsilon_\alpha)\psi_0& = & t_L \psi_1 +t_L \psi_{-1}  \nonumber \\
     (E-\epsilon_\alpha)\psi_1& = & t_L \psi_2 +t_L \psi_0 \\
     (E-\epsilon_\beta)\psi_2& = & t_S \psi_3 +t_L \psi_1  \nonumber \\
     (E-\epsilon_\gamma)\psi_3& = & t_L \psi_4 +t_S \psi_2 \nonumber \\
\vdots \nonumber
\end{eqnarray}

          If we generate a renormalized PD chain by decimating an
appropriate set of sites ( see Fig.1(a) ),then  we  obtain  a  self-similar
hierarchy of equations, but with renormalized  parameters corresponding  to
the inflated
chain shown in Fig.1(a).  The topology of  the  chain  is  preserved  as   a
result
of this transformation and  the  corresponding  scale  factor  is  two.  The
renormalized site energies and the hopping integrals  are found to be
\begin{eqnarray}
\epsilon_\alpha  '  &  =  &  \epsilon_\gamma  +\omega_\beta  (t_L^2  +t_S^2)
\nonumber
\\
\epsilon_\beta ' &  = &  \epsilon_\gamma  +\omega_\alpha  t_L^2+\omega_\beta
t_S^2 \nonumber \\
\epsilon_\gamma  '  &  =  &  \epsilon_\alpha   +(\omega_\alpha+\omega_\beta)
t_L^2\\
t_L ' & = & \omega_\beta t_L t_S \nonumber \\
t_S ' & = & \omega_\alpha t_L^2   \nonumber
\end{eqnarray}
     where      $\omega_i      =      1/(E      -      \epsilon_i)$,      $i
=
\alpha$, $\beta$, $\gamma$. Corresponding to this transformation it can   be
shown
that there exists the following polynomial invariant
\begin{equation}
     {\cal I}=\frac{(\epsilon_\alpha  -   \epsilon_\beta)   (\epsilon_\alpha
-
\epsilon_\gamma) - t_L^2 -t_S^2}{2 t_L t_S} + 1 ,
\end{equation}
               whose  value  remains  unchanged  under  the   transformation
Eq.(4),
as can be easily verified. Since  the  renormalized  chain  is  still  a  PD
sequence, we again find $\alpha$-sites occuring as $\alpha -\alpha  $  pairs
and $\beta -\gamma $ sites forming $\beta -\gamma  $  doublets.  This  means
that the $\alpha -\alpha $ clustering  effects  are  also  present  on  this
inflated length scale. With respect to this renormalized lattice  there  are
extended eigenstates at $E =\epsilon_\alpha '$ (when $M_\alpha^2 =  -I$),
provided   for   this   energy     $\kappa   '    =     (1/2)     |\mbox{\rm
Tr}M_{\beta
\gamma}|\leq 1$, $\alpha$, $\beta$ and $\gamma$ now referring to the
renormalized chain. Since  the $\alpha-\alpha$ clustering is present at
every length  scale, one will  find,  upon  repeated   renormalization,
a  number  of  extended
state-energy eigenvalues  by  solving  the  equation  $E  =  \epsilon_\alpha
^{(n)}$, and by checking  that  the  roots  of  the  equation  satisfy  the
condition $\kappa^{(n)} \leq 1$, where the trace $\kappa^{(n)}$ is evaluated
with the renormalized parameters at the $n$-th stage. Interestingly, for the
PD chain  we  find  that  the  value  of  $\kappa^{(n)}$  becomes  equal  to
${\cal I}-1$  for  energies  which  are  the  roots  of  the   equation   $E
-
\epsilon_\alpha^{(n)}=0$, where, ${\cal I}$  is  the  invariant  defined  in
Eq.(5).  This  implies  that   $\kappa^{(n)}$   is   also   independent   of
the
generation index $n$ for these special  energy  values.  Therefore,  if  the
initial parameters are such that $\kappa^{(0)}$ is greater  than  one,  then
there will be no extended states at any level of  renormalization,  and  the
roots of the equation $E-\epsilon_\alpha^{(n)}=0$ will always be in the gaps
of the entire band.  On the  other  hand,  if  for  the  initial  choice  of
parameters, $\kappa^{(0)}$ is less than or equal  to  unity,  then  all  the
subsequent $\kappa^{(n)}$'s  will  also  be  $\leq  1$,  and  thus  all  the
solutions
of the equation $E - \epsilon_\alpha^{(n)}(E)=0$ will be allowed ones.

           It is to be appreciated that the $\alpha -\alpha$  clustering  at
larger length scales amounts to including, so to say, the effects of  larger
and larger segments of the original chain into the $\alpha$-subclusters. The
$\alpha -\alpha$ correlation at all length scales can only  be  revealed  by
renormalization  group  methods.  While  in  the   original   lattice   this
correlation is directly visible (Fig.1(a)), higher  order  correlations  due
to
$\alpha -\alpha$  pairing  imply  underlying  complex  correlations  between
atoms,
which is not apparent from mere inspection of the original lattice.

               With this background we are now  in  a  position  to  discuss
several models of the PD chain which are  obtained  by  assigning  different
values to the hamiltonian parameters. Specifically,  we  shall  discuss  the
following models:

\vskip .5cm
\noindent
(i) \underline {The on-site model}

     In this model the hopping matrix elements are all taken to be equal and
the site energies are of two types, $\epsilon_A$ and $\epsilon_B$,  arranged
on a lattice following the PD  sequence.  Severin  et  al.$^{12}$  in  their
analysis showed that this  chain  supports  extended  states  which  display
periodicity of periods 4, 8 $\ldots$ etc. in units of lattice spacing.  They
made use of the fact that  a  pseudo-invariant
associated with this model has a value equal to 2, which is then used
to establish the periodic nature of the solutions. From our point  of  view,
we recover the on-site model by putting $\epsilon_\alpha = \epsilon_\gamma =
\epsilon_A$, $\epsilon_\beta =  \epsilon_B$  and  $t_L  =  t_S  =  t$.  For
numerical calculations we choose $\epsilon_A = 1$, $\epsilon_B = -1$,  $t  =
1$ in suitable units, which are the same as those used by Severin et al., in
order to afford a comparison with their calculation. At the basic level  the
extended state occurs at $E = \epsilon_A = 1$, and the wavefunction for this
energy is found by the transfer matrix procedure beginning  with  the  value
$\psi_1=1$ and $\psi_0=0$ for a 7th generation PD chain with 128 atoms.  The
wavefunction is shown in Fig.2(a) and is identical with  that   of   Severin
et
al.$^{12}$ having a periodicity of 4 units of lattice spacing.  This  energy
value yields for the quantity $\kappa^{(0)}$ the value -1, corresponding  to
the edge of the unperturbed $\beta \gamma $ band  of  energies.  Since  this
quantity       equals        ${\cal   I}        -1$,        the        trace
values
$\kappa^{(1)},\kappa^{(2)}$,\ldots $\kappa^{(1)}$ \ldots, are all  equal  to
-1. If we now consider the renormalized  chain,  then  the  extended  states
occur at energies obtained from the equation  $E  -\epsilon_\alpha  '(E)=0$,
the roots being $\pm \sqrt{3}$. The wavefunction for $E=\sqrt{3}$ is
shown in Fig.2(b), with a period of 8 units, and  is again  identical
with that of Severin et al.$^{12}$ At the next level we have  the  energies
by solving $E -\epsilon_\alpha ''(E)=0$, which yields the roots
$E$ = -1.82595501, -0.15244466, 1.38553731 and 2.59286237.  In
Fig.2(c) we give the wavefunction for $E$ =  2.59286237.  This  function  is
seen
to have a period of 16 units. Finally, in Fig.2(d) we plot the wavefunction
for  the  energy  $E$  =  2.60380559,  obtained  from  the  next  level   of
renormalization.

               As was pointed out by Severin et al.,  the  underlying
reason for the observed periodicity of say, the eigenfunction  corresponding
to $E = 1$ having a period of four  lattice  spacings,  is  related  to  the
vanishing of the amplitude at every fourth site on the chain. Alternatively,
we  note  that  at   this   energy,   the   hierarchy   of    Eqs.(3)    are
indistinguishable
from that of an infinite periodic lattice of alternate $A$  and  $B$  atoms.
Similarly, the equations determining the  eigenfunctions  of  period  8  are
identical with that of an ordered  lattice  with  unit  cell  consisting  of
$ABAA$ atoms. In the same manner it is found that every periodic solution in
the on-site model is identical with the solution of some periodic chain with
a suitable unit cell. Thus it turns out that in this case we may associate a
whole array of periodic lattices with different unit cells with the extended
eigenfunctions, each periodic lattice being in  one  to  one  correspondence
with one of the eigenvalues of the PD chain. A similar observation has been
made recently by Oh and Lee$^{16}$.

\vskip .5cm
\noindent
 (ii) \underline {The transfer model}

          As mentioned before, the transfer model is obtained by choosing the
starting values $\epsilon_\alpha  =\epsilon_\beta  =\epsilon_\gamma$,  and
$t_L \neq t_S$. For numerical work we choose  the  values  $\epsilon_\alpha
=\epsilon_\beta =\epsilon_\gamma =0$, $t_L =1$ and  $t_S  =  2$.  For   this
case
the energy for the extended state $E = 0$ lies in the  central  gap  of  the
band, with a value of $\kappa^{(0)} = -1.25$. Again, from the invariance of
the trace under the RSRG transformation, it  follows  that  $\kappa^{(n)}  =
-1.25$ for all $n$, that is, all the energy values are disallowed  and  thus
there are no extended states in this model. In fact, in this model the value
of $\kappa^{(n)}$ for  any  choice  of  $t_L$  and  $t_S$  is  of  the  form
$\kappa^{(n)}  =  (1/2)  (  x  +  1/x)$,  where,  $x  =t_L/t_S$,  and   thus
$\kappa^{(n)}$ is always greater than unity. Since the value of  ${\cal  I}$
in
this case is non-zero, the eigenstates  are  all  critical  supported  on  a
Cantor set of zero measure.

\vskip .5cm
\noindent
 (iii) \underline {The mixed model}

     Let us begin with the case where all the site energies are unequal, and
there are two  different  hopping  matrix  elements  $t_L$  and  $t_S$.  The
transfer matrix $M_\alpha$ for  this  case  is  unimodular,  and  hence  the
energies for the extended states are again obtained  from  the  equation  $E
-\epsilon_\alpha^{(n)} (E)=0$. The band of energies allowed by the periodic
$\beta \gamma$  chain defined by the initial parameters lies  in  the  range
$$
\left ( \frac {\epsilon_\beta + \epsilon_\gamma + \sqrt{(\epsilon_\beta -
\epsilon_\gamma )^2 + 4 (t_S + t_L)^2}}{2}, ~
  \frac {\epsilon_\beta + \epsilon_\gamma + \sqrt{(\epsilon_\beta -
\epsilon_\gamma )^2 + 4 (t_S - t_L)^2}}{2} \right )
$$
\noindent
and
$$
\left ( \frac {\epsilon_\beta + \epsilon_\gamma - \sqrt{(\epsilon_\beta -
\epsilon_\gamma )^2 + 4 (t_S - t_L)^2}}{2}, ~
  \frac {\epsilon_\beta + \epsilon_\gamma - \sqrt{(\epsilon_\beta -
\epsilon_\gamma )^2 + 4 (t_S + t_L)^2}}{2} \right )
$$
If the value of  $\epsilon_\alpha$   lies  outside
these  energy intervals, then $\kappa^{(0)}$ becomes greater than  unity  at
the energy $E = \epsilon_\alpha$ and there is no extended state at any level
of renormalization. With the progress of renormalization the periodic $\beta
\gamma$ chains with  renormalized  parameters  will   yield  more  and  more
fragmented bands,  and  the  energy  values  obtained  as  solutions  of  $E
-\epsilon_\alpha^{(n)} (E)=0$ will always be in the gaps of these fragmented
spectrum. Since the value of ${\cal I}$ is non-zero, all  the  states   will
be
critical even in the presence of  $\alpha   -\alpha$  clustering.   On   the
other
hand, if the value of $\epsilon_\alpha$  lies  inside  one  of  the  initial
energy intervals defined above, we always have $\kappa^{(n)} \leq 1$,  and
the states will be extended. The extended character of each wavefunction has
been checked by a mutifractal analysis (see section IV).  There
is thus a cross-over in the behaviour of the eigenstates  depending  on  the
choice of the initial parameters. Confining our attention to the  regime  of
extended behaviour, we find an interesting systematics in the nature of  the
eigenfunctions as we consider the roots of  $E  -\epsilon_\alpha^{(n)}(E)=0$
for  increasing  values  of   $n$. Choosing $\epsilon_\alpha = 4$,
$\epsilon_\beta = 0$ and $\epsilon_\gamma = 3$, $t_L = 1$ and $t_S = 2$,
we find that all   the   wavefunctions   are
non-periodic, as can be seen from Fig.(3), where we  have  displayed  the
eigenfunctions at the energies $4$, $-1.19258240$, $5.45570651$ and
$3.56527726$ arising from the
first four levels of renormalization. In  Fig.(4)  eigenfunctions
are shown for energies $-0.60581006$, $4.14562305$, $4.65276530$ and
$5.45820519$, which arise  from  the  solution
of  $E  -\epsilon_\alpha^{(4)}(E)=0$.
As may be seen, the amplitudes of these non-periodic functions do not  decay
as we go  from one end of the  chain  to  the  other.  We  have  tested  the
non-decaying character of the wavefunctions for chain lengths upto $2^{18}$
atoms,
although we have presented the amplitudes for much smaller chain lengths for
convenience.  A  very  interesting  feature  that  may  be  noted  is   that
eigenfunctions corresponding to neighbouring energies possess,  in  general,
entirely different profiles. This feature is contrary to what one obtains in
a periodic lattice, and one may regard this behaviour as a manifestation  of
the quasiperiodic character of the lattice.

               We now go on to the discussion of a still more general model,
namely one in which, in addition to having three different site-energies, we
ascribe four different values to the hopping integral across the long  bonds
connecting $\alpha \alpha$, $\alpha \beta$,  $\gamma  \alpha$  and  $\gamma
\beta$ pairs of sites. While the primary reason for considering  this  model
is that it represents  the  realistic  situation  most  closely  within  the
framework of the single band nearest neighbour  tight-binding  hamiltonian,
an additional reason is that even though $\alpha-\alpha$ clusters  are  seen
to be present,
the matrix $M_\alpha$ will  have  two  different  values  depending  on  the
immediate neighbours of an $\alpha$ site. Neither of  these  two  $M_\alpha$
matrices is unimodular, and  therefore  one  cannot  apply  the  result   of
Eq.(2)
to this case. Inspite of this,
the present model gives  rise  to  an  infinity  of
extended states. To see this,  if we consider only the  subset  of  $\alpha$
sites in this PD chain, then these sites are found  to  form  a  lattice  in
which  two  different  bond  lengths  are  arranged   in   the   copper-mean
sequence$^5$. In the same manner, each of the subset of $\beta$ and $\gamma$
sites forms a copper-mean sequence. If we now decimate the original PD chain
so as to retain only the $\beta$ sites (the same could have been  done  for
the $\alpha$ or $\gamma$ sites), then we  may  relabel  the  sites  of  the
resulting lattice  following  the  prescription  of  the  copper-mean  chain
discussed by us earlier$^5$, so that the original PD chain transforms into a
copper mean lattice (see Fig.1(b)) with four site-energies
$\epsilon_\alpha^{CM}$,
$\epsilon_\beta^{CM}$, $\epsilon_\gamma^{CM}$ and  $\epsilon_\delta^{CM}$
and two
different hopping integrals $t_L^{CM}$ and $t_S^{CM}$ defined as

\begin{eqnarray}
\epsilon_\alpha^{CM}& = &  \epsilon_\beta  +  s  t_S^2 \omega_\gamma  +
q t_{\alpha\beta}^2 \omega_\alpha \nonumber \\
\epsilon_\beta^{CM}& = & \epsilon_\beta + t_S^2 \omega_\gamma
+ q t_{\alpha\beta}^2 \omega_\alpha  \nonumber \\
\epsilon_\gamma^{CM}& = &  \epsilon_\beta  +  s  t_S^2 \omega_\gamma  +
t_{\gamma\alpha}^2 \omega_\gamma \\
\epsilon_\delta^{CM}&  =  &  \epsilon_\beta  +  t_S^2 \omega_\gamma  +
t_{\gamma\alpha}^2 \omega_\gamma \nonumber \\
t_L^{CM}   &   =   &   p~ q~ t_S~  t_{\gamma \alpha}   t_{\alpha\alpha}
t_{\alpha\beta} \omega_\gamma {\omega_\alpha}^2 \nonumber \\
t_S^{CM} & = & t_{\gamma\alpha}~ t_S \omega_\gamma ,     \nonumber
\end{eqnarray}
        where $\omega_i$'s have been defined previously. $p,~q,~r$ and $s$
are given by
\begin{eqnarray}
p & = & (1-t_{\gamma \alpha}^2 \omega_\alpha \omega_\gamma )^{-1}  \nonumber
\\
q & = & (1-p t_{\alpha \alpha}^2 \omega_\alpha^2  )^{-1} \nonumber \\
r & = & (1- t_{\alpha \alpha}^2 \omega_\alpha^2 )^{-1} \nonumber \\
s &  =  &  (1-r  t_{\gamma  \alpha}^2  \omega_\alpha  \omega_\gamma  )^{-1}.
\nonumber
\end{eqnarray}

               Very  interestingly  we  find  that  as  a  result  of   this
transformation the hopping matrix elements  for  the  effective  copper-mean
chain takes only two different values depending on the two distinct bonds in
this renormalized lattice, even though we had started with the most  general
version of the PD chain. In this renormalized copper-mean  lattice  we  have
again the $\alpha -\alpha $ clustering as discussed in Ref.$5$, and this
leads to a whole hierarchy of extended states,  provided  $(1/2)  |\mbox{\rm
Tr}
(M_\gamma M_\delta M_\beta)| \leq 1$.

               Two comments are in order  at  this  stage.  Firstly,  it  is
impossible to discern this clustering effect at the level of the original PD
chain, and the full spectrum of extended states can be obtained only through
this initial transformation to a copper-mean lattice. This means,  that  the
original PD chain does implicitly contain a complex clustering involving all
the $\alpha$, $\beta$ and $\gamma$ sites, but which does not show up at  the
initial stage. Secondly, it may  seem  that  one  could  have   straightaway
applied this transformation to any of the models discussed earlier. However,
it should be appreciated that there is a basic difference between  the  RSRG
transformation for a PD lattice and that  for  a  copper-mean  chain.  While
there exists a polynomial invariant ${\cal I}$ associated with the  PD
chain,  there
is no such invariant for the copper-mean case. The  value   of   ${\cal  I}$
for
the most general model is found to be
\begin{equation}
{\cal I}  =   \frac{(E     (t_{\alpha\alpha}^2
-t_{\gamma\beta}^2)      -
t_{\alpha\alpha}^2  \epsilon_\gamma  +  \epsilon_\alpha  t_{\gamma\beta}^2)
(\epsilon_\alpha  -\epsilon_\beta)  -   t_S   t_{\alpha\alpha}^2   (t_S   -2
t_{\gamma\beta})    -    t_{\alpha\alpha}^2    t_{\gamma\beta}^2}{2      t_S
t_{\gamma\beta}
t_{\alpha\alpha}^2}.
\end{equation}

        In determining ${\cal I}$ we have set the initial hopping integrals
$t_{\alpha \alpha} = t_{\alpha \beta }$ and $t_{\gamma \alpha}  =  t_{\gamma
\beta }$.
This is, of course, not a restrictive condition,  since  starting  with  the
most
general model with four different values for the hopping integral for the
long bonds, we get, upon decimating alternate sites following the PD
renormalization rule, only two different values for this hopping ,which we
may choose to be $t_{\alpha\alpha}$ and $t_{\gamma\beta}$.
The quantity ${\cal I}$ is a linear function of energy, in contrast  to  the
any
of
the other models discussed earlier. By setting ${\cal I} = 0$ (see Ref.7 and
8),
we
immediately find that there is an additional extended state at an energy
\begin{equation}
E  = \frac{(\epsilon_\alpha  -  \epsilon_\beta  )(\epsilon_\gamma  t_{\alpha
\alpha}^2
    - \epsilon_\alpha t_{\gamma \beta }^2) + t_{\alpha \alpha }^2 (t_S -
    t_{\gamma  \beta})^2}{(\epsilon_\alpha  -  \epsilon_\beta  )  (t_{\alpha
\alpha}^2
    - t_{\gamma \beta}^2)}.
\end{equation}

 We have computed this wavefunction numerically with $\epsilon_\alpha = 1$,
  $\epsilon_\beta = 0$, $\epsilon_\gamma = 0.1$, $t_{\alpha \alpha } = 1$,
  $t_{\gamma \alpha } = 1.1$ and $t_S = 2$, for which $E$ is found to have a
  value $1.428571428$. This extended state  cannot  be  detected   from  a
  consideration  of  the
effective copper mean lattice. The  character  of  this  extended  state  is
entirely different from the other extended states in the PD lattice, if  one
looks  at  the  pattern  of  flow  of  the  hamiltonian  parameters  in  the
copper-mean chain with renormalization.  As  has  been  pointed  out  by  us
earlier$^5$,   for   an   energy   obtained    from    the    equation    $E
-\epsilon_\alpha^{(n)} (E) = 0$ for the effective copper-mean  lattice,  the
values of  $\epsilon_\beta$  and  $\epsilon_\gamma$  become  equal  after  a
certain number of iterations depending on the value of $n$, and remain so in
subsequent iterations. Moreover, the hopping integrals do not flow to  zero,
which is an evidence of  the extended nature of these eigenfunctions. On the
other  hand,   for   the   special   energy   value   defined   in   Eq.(8),
$\epsilon_\beta$
is never  found  to  become  equal  to  $\epsilon_\gamma$   at   any   stage
of
iteration, although the hopping integrals do  remain  finite  and  non-zero,
indicating that this state is also extended. The chaotic  behaviour  of  the
flow pattern may be said to characterise this eigenstate. The  corresponding
wavefunction is shown in Fig.(5), and its extended character is confirmed
through the multifractal analysis given  in section (IV).

\vskip .5cm
\noindent
{\bf    III. The Multi-Band Model}

          Let us now consider  the  continuum  version  of  the   Scrodinger
equation in one dimension with a quasi-periodic  potential  $V(x)$.  We  may
transform this  equation with any arbitrary  potential to a discrete set  of
difference equations using the standard Poincare map$^{13}$. If we label the
points on a $1d$ lattice by  the  integers  ${x_i}$,  then  we  can  find  a
recursion  relation  for  the  amplitudes  of  the  wavefunction  at   three
consecutive points of the above set. In the usual language of tight  binding
theory, the ``hopping matrix element" connecting amplitudes on neighbouring
points $x_i$ can be expressed in terms of the Wronskian matrix corresponding
to the solutions of the Schrodinger equation in that interval. We  omit  the
expressions here  and  refer  to  the  literature  for  details$^{13}$.  For
simplicity,  let  us  apply  this  theory  to  the  case  of  an  array   of
$\delta$-function potentials sitting at the sites  $i$  with  strengths
$\lambda_i$ distributed according to the PD sequence.The Poincare map  leads
us to a set of difference equations of the following form connecting the
amplitudes at three successive points:

\begin{equation}
\psi_{n+1} +\psi_{n-1}  = [ 2 \cos q + (\lambda_n /q) \sin q ] \psi_n
\end{equation}
where $q = \sqrt(E)$. Here $\lambda_n$ gives the  strength  of  the
potential at the $n$-th  site.  The  energy  eigenvalues  forming  bands  of
extended states correspond to the values of the wave vector  $q$  satisfying
the inequality
\begin{equation}
2 \cos q + (\lambda_n /q) \sin q \leq 2 ~.
\end{equation}
In order to extract information about the spectral nature  as  well  as  the
character of the eigenfunctions in this case we can recast Eq.(9)
in the following form :
\begin{equation}
(E '- \epsilon_n )\psi_n  =  \psi_{n-1} + \psi_{n+1}~,
\end{equation}
with $E' = 2 \cos q$ and $\epsilon_n = -(\lambda_n \sin q)/q$.
This is now in the  familiar
form of the single band tight binding hamiltonian where $E '$ now plays  the
role of the `energy' and $\epsilon_n$  can  be
interpreted as the `effective' on site potential  distributed  according  to
the PD sequence. We may now assign three different labels $\alpha$, $\beta$
 and $\gamma$ to the sites of the chain and adopt the same procedure as
 described earlier for finding  the  extended  eigenstates.
Setting $E'=\epsilon_\alpha^{(n)}$, where, $n$ implies the $n$-th  stage  of
renormalization, we have been able to locate the  eigenvalues  corresponding
to the extended eigenfunctions for the multiband PD chain. There is however,
an important aspect to note, viz. setting $E '=\epsilon_\alpha$ even at  the
initial stage we arrive at an equation of the form
\begin{equation}
2/\lambda_\alpha  = - (1/q) \tan q.
\end{equation}
Now $\tan q$ is a $\pi$ -periodic function and it takes all values in  the
interval $-\infty$ to $\infty$ for any value of $\lambda_\alpha$. Therefore,
at  each
stage   of   renormalization   the   solution    of    the    equation    $E
'=\epsilon_\alpha^{(n)}$ will give rise to  an  infinite  countable  set  of
energy eigenvalues for which the PD chain will be totally transmitting. That
is to say, we have infinite number of extended eigenfunctions at all  scales
of length. This result is to be contrasted with the case of a tight  binding
random dimer model which exhibits a resonance state at a unique  energy,  or
with the results for the copper-mean chain or the PD chains in the tight
binding  versions
where, at each stage of renormalization one generally  enumerates  a  finite
number  of  extended  eigenstates.  For   numerical   calculation   of   the
wavefunctions
 we have taken $\lambda_\alpha = \lambda_\gamma = 1$  and  $\lambda_\beta  =
-1$.
 At the very initial stage, by  setting  $E  '=\epsilon_\alpha$  we
obtain an  infinite  number  of  allowed  $q$  values.  We  have  explicitly
calculated the wavefunction for $q = 1.836597203$ and the result is shown in
Fig.6(a). This wavefunction is real and  periodic  with  a  period
four. The amplitudes follow the sequence $1, 0, -1, 0$ for the first
four sites and this pattern is repeated for the rest  of  the  lattice.  The
discontinuity of the  derivative  of  the  wavefunction  at  lattice  points
$x=3na$, $n = 1, 2, 3...$ have been shown encircled in Fig.6(a). From the
next level of renormalization, we have chosen $q = 2.37738316$ from the set
of allowed $q$- values, and have plotted the wavefunction, which has a
periodicity of eight, in Fig.6(b). The wavefunctions for other levels of
renormalization may be obtained in the same manner.

\vskip .5cm
\noindent
{\bf IV. Multifractal Analysis of the Wavefunctions}

          All the wavefunctions given in this paper have been  subjected  to
a multifractal analysis  to  verify   their   extended   nature.   Following
the
standard procedure, a ``partition function" is defined as
\begin{equation}
Z(Q)  =  \sum_{i=1}^{N}  |\psi_i|^Q
\end{equation}
where $N$ is the number of sample points. For sufficiently large $N$, $Z(Q)$
may be expressed as
\begin{equation}
Z(Q) = \epsilon^{\tau(Q)}
\end{equation}
where $\epsilon$ is the interval $1/N$, and $\tau(Q)$ is an index from which
the multifractal index $\alpha$ and the corresponding fractal dimension
$f(\alpha)$
are obtained as
\begin{equation}
\alpha  = \frac{d \tau(Q)}{d Q}~~~\mbox{\rm and}~~~
f(\alpha)  =  Q \alpha -\tau(Q) ~~.
\end{equation}

 As is known$^{15}$,  the  standard  algorithm  is  however  not  of  much
practical utility unless the sample size is  inordinately  large.  With  all
finite, reasonably large, sample sizes one always obtains a distribution  in
the values of $\alpha$ and $f(\alpha)$. This is a finite-size effect and the
converegence to the true behaviour occurs only logarithmically  slowly  with
increase in the sample size$^{15}$. Thus the standard  procedure  cannot  be
regarded
as a reliable method to analyse finite size samples. As has been  explicitly
shown by Godreche and Luck$^{15}$,  a  much  better  method  to  test  the
extended character is to look at the behaviour of the curvature of $\alpha -
f(\alpha)$ curve at $Q = 0$.  As  has  been  shown  by  them,  the  quantity
$|1/f ''(\alpha)|_{Q=0}$, which is a measure  of  the  curvature at the
maximum of the $\alpha -f(\alpha)$ graph,
should diverege linearly with $\log(N)$ for  extended  states,   $N$   being
the
sample size. On the contrary, if the given set is a true  multifractal,  the
curvature should gradually saturate to single  value  with   $\log(N)$.   In
our
case we have used this  prescription  to  test  our  wavefunctions,  with  a
maximum sample size of $2^{17}$ values of the amplitudes. In every case, we
observe the  predicted  linearity  in  the  dependence  of
$|1/f''(\alpha)|_{Q=0}$
with $\log(N)$. In Fig.7 we give the results for three  of
the wavefunctions, including the `chaotic' wavefunction shown in Fig.5, the
graphs for all the other wavefunctions being entirely similar. Our analysis,
therefore provides confirmatory evidence that the  states  obtained  on  the
basis of the present work are true extended states.

\newpage
\noindent
{\bf References}
\begin{itemize}
\item[$^1$] D. Shechtman, I. Blech, D. Gratias and J. W. Cahn,
Phys. Rev. Lett. {\bf 53},
1951 (1984).
\item[$^2$] R. Merlin, K. Bajema, R. Clarke, F. Y. Juang and
P.  K.  Bhattacharya, Phys. Rev. Lett. {\bf 55}, 1768 (1985).
\item[$^3$] M. Kohmoto. B. Sutherland and C. Tang,
Phys. Rev. {\bf B 35}, 1020 (1987);
  J. M. Luck and D. Petritis, J. Stat. Phys. {\bf 42}, 289 (1986);
  R. Riklund  and M. Severin, J. Phys. {\bf C 21}, L965 (1988).
\item[$^4$] M. Kohmoto, L. P. Kadanoff and C.  Tang,
Phys.  Rev.  Lett.  {\bf 50},  1870 (1983).
\item[$^5$] S. Sil, S. N. Karmakar, R. K. Moitra  and
Arunava  Chakrabarti,  Phys. Rev.{\bf B 48}, 4192 (1993).
\item[$^6$] D. H. Dunlap, H. L. Wu and P. Phillips,
Phys. Rev. Lett. {\bf 65}, 88 (1990).
\item[$^7$] V. Kumar and G. Ananthakrishna,
Phys. Rev. Lett. {\bf 59}, 1476 (1987);
\item[$^8$] Arunava Chakrabarti, S. N. Karmakar and R. K.  Moitra,
Phys.  Lett. {\bf A 168}, 301 (1992);
Arunava Chakrabarti, S. N. Karmakar and R. K. Moitra,
Phys. Rev. {\bf B 39}, 9730 (1989).
\item[$^9$] R. Riklund, M. Severin and Y. Liu,
Int. J. Mod. Phys. {\bf B 1}, 121 (1987).
\item[$^{10}$] J. M. Luck, Phys. Rev. {\bf B 39}, 5834 (1989).
\item[$^{11}$] J. Bellissard, A. Bovier and J. M. Ghez,
Commun.  Math.  Phys. {\bf 135}, 379 (1991).
\item[$^{12}$] M. Severin, M. Dulea and R. Riklund,
J. Phys. Cond.  Matt. {\bf 1},  8851 (1989).
\item[$^{13}$] J. Bellissard, A. Formoso, R. Lima and D. Testard,
Phys. Rev. {\bf B  26}, 3024 (1982);
D. Wurtz, M. P. Sorensen and T. Schneider, Helv. Phys. Acta {\bf 61},
363   (1986); A. Sanchez, E. Macia and F. Dominguez-Adame,
Phys. Rev. {\bf B  49},147 (1994).
\item[$^{14}$] B. W. Southern. A. A. Kumar, P. D. Loly
and A. M. S. Tremblay, Phys. Rev.  {\bf B 27}, 1405 (1983).
\item[$^{15}$] C. Godreche and J. M. Luck,
J. Phys. A: Math. Gen. {\bf 23}, 3769  (1990).
\item[$^{16}$] G. Y. Oh and M. H. Lee, Phys. Rev. {\bf B 48}, 12465 (1993).

\end{itemize}
\newpage
\begin{center}
{\bf FIGURE CAPTION}
\end{center}
\begin{itemize}
\item[Fig.1] (a) A section  of  the  period-doubling  chain.  The  $\alpha$,
$\beta$
and $\gamma$ sites are indicated respectively by a full circle, a square and
an open circle. $\alpha - \alpha$ clusters are shown encircled by dotted
lines. The renormalized chain under a period-doubling decimation is also
shown.
(b) Effective copper-mean chain obtained by eliminating the $\alpha$ and
$\gamma$ sites in the period-doubling chain. The $\alpha$, $\beta$, $\gamma$
and $\delta$ sites are marked by a full circle, square, open circle and a
triangle in the copper-mean chain. $\alpha - \alpha$ clusters in the copper-
mean chain are also indicated.
\item[Fig.2] Wavefunctions for the period-doubling chain in the on-site
model for $\epsilon_\alpha = \epsilon_\gamma =1$, $\epsilon_\beta = -1$,
$t_L = 1$ and $t_S = 2$. For (a) $E = 1$, (b) $E = \sqrt{3}$, (c) $E =
2.59286237$ and (d) $E = 2.60380559$. All energies are in units of $t_L$.
\item[Fig.3] Wavefunctions for the period-doubling chain in the mixed
model for $\epsilon_\alpha = 4$, $\epsilon_\beta =0$, $\epsilon_\gamma = 3$,
$t_L = 1$ and $t_S = 2$. For (a) $E = 4$, (b) $E = -1.19258240$, (c) $E =
5.45570651$ and (d) $E = 3.56527726$. All energies are in units of $t_L$.
\item[Fig.4] Wavefunctions for the period-doubling chain in the mixed
model. Parameters are the same as for Fig.(3). The energies are for
(a) $E = -0.60581006$, (b) $E = 4.14562305$, (c) $E = 4.65276530$ and
(d) $E = 5.45820519$.
\item[Fig.5] Wavefunction for the general model :
$\epsilon_\alpha = 1$, $\epsilon_\beta = 0$, $\epsilon_\gamma = 0.1$,
$t_{\alpha \alpha} = t_{\alpha \beta} = 1$ and $t_{\gamma \alpha} =
t_{\gamma \beta} = 1.1$, and $t_S = 2$, for $E = 1.428571428$ obtained from
Eq.(8). The energies are in units of $t_{\alpha \alpha}$.
\item[Fig.6] Wavefunctions for the period-doubling chain in the multi-band
case for $\lambda_\alpha = \lambda_\gamma =1$, $\lambda_\beta = -1$
for (a) $q = 1.836597203$ corresponding to a period four and (b)
$q = 2.377380316$ for a period eight. The energies are measured in
units such that $\hbar^2/2m =1$. The discontinuities in $\psi '(x)$ are
shown encircled.
\item[Fig.7] Plot of $|1/f''(\alpha)|_{Q=0}$ (curvature) against $\log(N)$
corresponding to  wavefunctions in Fig.2(d) , Fig.3(d) and Fig.5 are shown
in (a), (b) and (c) respectively. The maximum sample size has been taken as
$2^{17}$.
\end{itemize}
\end{document}